\documentclass[10pt]{article}

\usepackage{amsmath}
\usepackage{amssymb}

\usepackage{graphicx}

\usepackage{cite}
\usepackage{color} 

\usepackage{setspace} 
\usepackage[labelfont=bf,labelsep=period,justification=raggedright]{caption}

\makeatletter
\renewcommand{\@biblabel}[1]{\quad#1.}
\makeatother

\date{}

\usepackage{pdfcomment}

\begin{document}

\begin{flushleft}
{\Large
\textbf{A reaction-diffusion model of cholinergic retinal waves}
}
\\
Benjamin Lansdell$^{1,\ast}$, 
Kevin Ford$^{2}$, 
J Nathan Kutz$^{1}$
\\
\bf{1} Department of Applied Mathematics, University of Washington, Seattle, Washington, USA
\\
\bf{2} Department of Biochemistry and Biophysics, University of California San Francisco, San Francisco, California, USA
\\
$\ast$ E-mail: Corresponding lansdell@uw.edu
\end{flushleft}

\section*{Abstract}


Prior to receiving visual stimuli, spontaneous, correlated activity in the retina, called retinal waves, drives activity-dependent developmental programs. Early-stage waves mediated by acetylcholine (ACh) manifest as slow, spreading bursts of action potentials. They are believed to be initiated by the spontaneous firing of Starburst Amacrine Cells (SACs), whose dense, recurrent connectivity then propagates this activity laterally. Their inter-wave interval and shifting wave boundaries are the result of the slow after-hyperpolarization of the SACs creating an evolving mosaic of recruitable and refractory cells, which can and cannot participate in waves, respectively. Recent evidence suggests that cholinergic waves may be modulated by the extracellular concentration of ACh. Here, we construct a simplified, biophysically consistent, reaction-diffusion model of cholinergic retinal waves capable of recapitulating wave dynamics observed in mice retina recordings. The dense, recurrent connectivity of SACs is modeled 
through local, excitatory coupling occurring via the volume release and diffusion of ACh. In addition to simulation, we are thus able to use non-linear wave theory to connect wave features to underlying physiological parameters, making the model useful in determining appropriate pharmacological manipulations to experimentally produce waves of a prescribed spatiotemporal character. The model is used to determine how ACh mediated connectivity may modulate wave activity, and how parameters such as the spontaneous activation rate and sAHP refractory period contribute to critical wave size variability.

\section*{Author Summary}

Both within the visual system and more generally, two general processes describe
nervous system development: first, genetically determined cues provide
a coarse layout of cells and connections and second, neuronal activity
removes unwanted cells and refines connections. This activity occurs
not just through external stimulation, but also through correlated,
spontaneously generated bursts of action potentials occurring in hyper-excitable
regions of the developing nervous system prior to external stimulation.
Spontaneous activity has been implicated in the maturation of many neural
circuits, however exactly which features are
important for this purpose is largely unknown. 

In order to help address this question we construct a mathematical model to understand
the spatiotemporal patterns of spontaneously driven activity in the developing
retina. This activity is known as retinal waves. We describe a simplified, biophysically
consistent, reaction-diffusion model of cholinergic retinal waves
capable of recapitulating wave dynamics observed in mice retina recordings.
This novel reaction-diffusion formulation allows us to connect
wave features to underlying physiological parameters.
In particular this approach is used to determine which features of the system are
responsible for wave propagation and for the spatiotemporal patterns of propagating waves observed in both
mice and other species.

\section*{Introduction}

Throughout the nervous system, correlated spontaneous activity drives
developmental programs \cite{Moody2005,Blankenship2010}. Within the
retina, these events manifest as slow, spreading waves of depolarizations
and are thus termed retinal waves. Waves are observed in a variety
of species and progress through three stages of development
\cite{Wong1999a,Ford2012a}. They have been implicated in numerous
developmental processes, including retinotopic map refinement \cite{Huberman2008}
and the eye-specific segregation of retinal projections into layers
of the thalamus \cite{Xu2011}. In mammals, waves mediated by acetylcholine (cholinergic waves, or stage
II waves) are the best characterized. They exhibit a slow wavefront velocity, random
initiation site and direction of propagation, an interwave
interval (IWI) which lasts tens of seconds, and constantly shifting wave boundaries. Precisely
determining their role in development requires a controlled manipulation
of these properties, which in turn requires a sound theoretical understanding
of the mechanisms responsible for their generation. However, the complexity
of their dynamics means that determining the connection between the generation and resulting spatiotemporal patterns of wave
activity and underlying physiology is by no means intuitive and
is reliant on computational modeling. This paper develops a biophysically
consistent, yet simplified, conductance based model of the developing retina that is able to produce physiological waves, in order to elucidate this connection.

Retinal waves are believed to be initiated by spontaneous depolarizations
of Starburst amacrine cells (SACs) whose processes reside in the inner plexiform layer (IPL)
of the retina. In mice these spontaneous depolarizations are sparse:
they occur roughly once every 15 minutes per SAC \cite{Ford2012}, where as in rabbit the rate is closer to once every 30 seconds. Following initiation, the dense, recurrent connectivity
of the SAC network laterally propagates activity through cholinergic
synapses \cite{Zheng2004}. After depolarizing, SACs exhibit a slow
after-hyperpolarization current due to a cyclic-AMP sensitive, calcium-activated
potassium channel \cite{Ford2012,Zheng2006}, which persists for tens
of seconds. Consequently, waves propagate over only a finite domain
of the retina, with their boundaries determined by regions still refractory
from previous wave activity. A number of computational studies have
tested this wave generation hypothesis, notably the models by Godfrey
\emph{et al} 2007 \cite{Godfrey2007}, Hennig \emph{et al} 2009 \cite{Hennig2009}
and subsequent study by Ford \emph{et al }2012 \cite{Ford2012} (refer
to the reviews Gjorgjieva and Eglen 2011 \cite{Gjorgjieva2011} and Godfrey and Eglen 2009 \cite{Godfrey2009}
for more information on previous computational studies).

A striking feature of retinal waves is their variety of sizes, speeds
and shapes, which differs from the more stereotyped behavior of
spontaneous activity in other developing brain regions, for example
in neocortex \cite{Adelsberger2005,Allene2010}. Despite numerous theoretical studies, it is not clear
what properties of the recurrent developing SAC network most contribute
to these spatiotemporal patterns. Indeed, retinal waves sizes appear to be distributed
according to a power-law\cite{Hennig2009}, analogous to other examples
of spontaneous activity in the nervous system. In cortex, such activity
has been extensively studied as an example of a critical state phase-transition
\cite{Beggs2003}. Additionally, diffuse release of ACh
has been detected at the inner limiting membrane coincident with wave
activity \cite{Ford2012}, suggesting that, like later stage waves
mediated by glutamate \cite{Blankenship2009}, extra-synaptic diffusion
may play a role in cholinergic wave propagation, though this has not
been tested theoretically or experimentally. 

In this work, we develop a theoretical, reaction-diffusion framework
that integrates the key biophysical processes, including increased
excitation due to acetylcholine diffusion and the slow after hyper-polarization
of the SACs, to characterize spontaneous wave dynamics in the developing
retina. This provides a framework to address the above questions of critical dynamics and mechanisms of wave generation and lateral propagation.
The model is consistent with the above described mechanisms, but its
purpose is to be as simple as possible while still being able to capture
the defining spatiotemporal properties of retinal waves. We demonstrate the model is capable of producing physiological waves, as observed in mice \cite{Ford2012}, after which
we address two questions. First, what are the conditions for spontaneous activity in the amacrine
cell layer to propagate laterally as waves? We derive necessary conditions for propagating activity using techniques from non-linear dynamics that are made available through the use of a reaction-diffusion model. These conditions are shown to be consistent with known pharmacological and genetic manipulations. Second, once conditions for propagation are met, what are the conditions required for the spatiotemporal patterns 
of retinal wave activity to take the form of avalanches, as observed in multielectrode array recordings of stage II waves in a variety of species \cite{Hennig2009}?  This is answered by appeal to a model of critical behaviour in a canonical model of forest fires.

\section*{Results}

\subsection*{A physiological reaction-diffusion model of cholinergic retinal waves}

\begin{figure}[!ht]
\begin{center}
\includegraphics{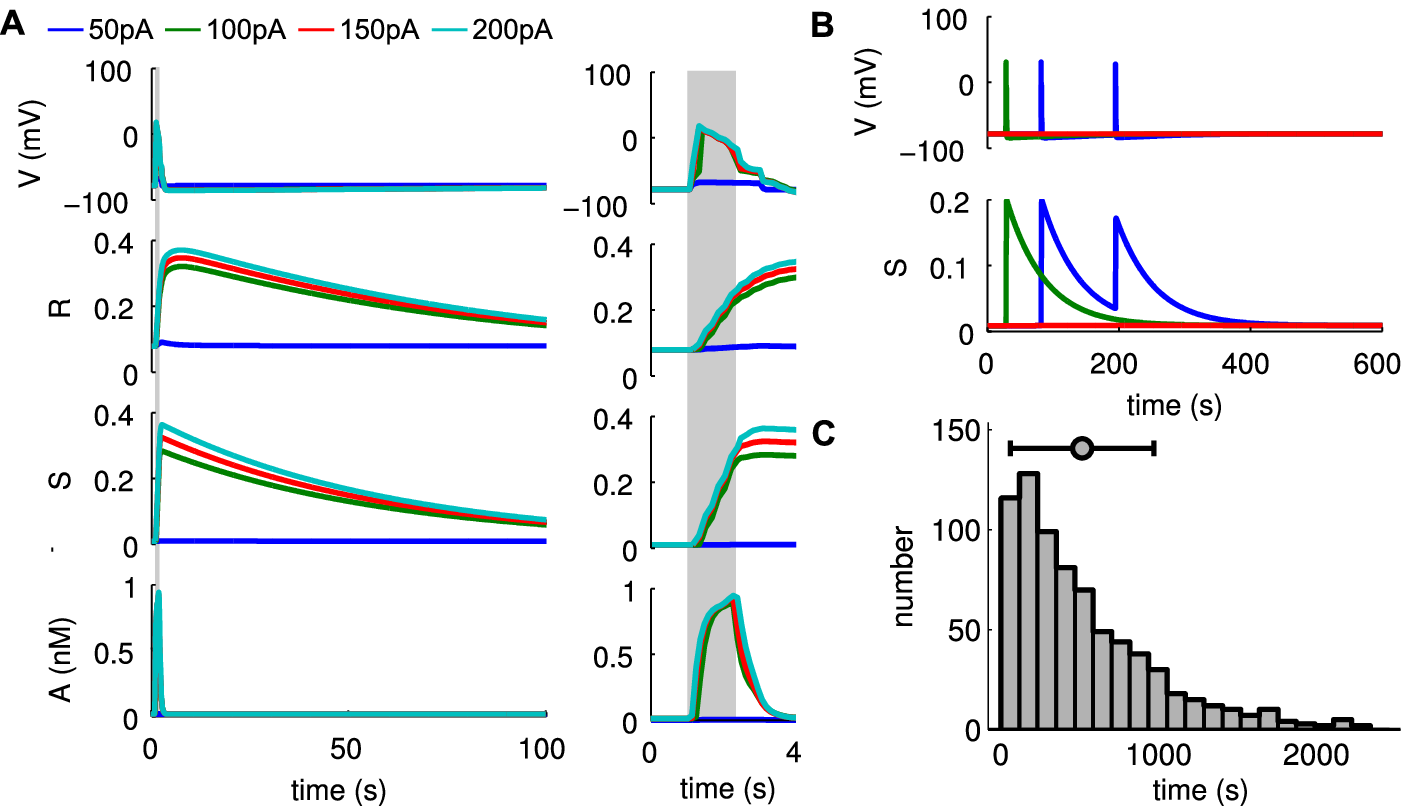}
\end{center}
\caption{\textbf{Isolated SAC dynamics}. Starburst amacrine
cells obey Morris-Lecar dynamics with voltage term $V$, refractory
variable $R$, sAHP variable \textbf{$S$ }and acetylcholine concentration
$A$. \textbf{A.} Time course of individual SAC dynamics following
current injections of indicated size, injected at $t=1\mbox{s}$ for
1.5s (shown in grey). Model SACs' refractory period shows dependence
on amount of current injected.\textbf{ B. }Single cell dynamics showing
sparse spontaneous depolarizations. Different colors represent different SACs. \textbf{ C. }Distribution of inter-event
intervals in stochastic simulation of isolated SAC. A threshold of
-50 mV was employed. Bar plot shows mean and standard deviation.}
\label{fig:isolatedsacs}
\end{figure}

As in previous models \cite{Hennig2009,Ford2012}, individual SACs
are modeled according to Morris-Lecar dynamics \cite{Morris1981a},
with an additional sAHP current activated by depolarization and subsequent
calcium influx. The sAHP current generates recovery times on the order
of a minute. Each cell is assigned the same recovery time scale, and
the dynamics are such that recovery is minimally activity-dependent
(Figure \ref{fig:isolatedsacs}A). Previous theoretical
models \cite{Godfrey2007,Hennig2009} and experimental observations
\cite{Zheng2006} show that larger depolarizations increase
sAHP duration. The lack of a strongly activity-dependent refractory
period in our  model does not affect the its ability to generate physiological
waves. A noisy, excitatory current induces spontaneous depolarizations
(Figure \ref{fig:isolatedsacs}B) at an average rate of once every
15 minutes per isolated cell, consistent with recordings in mice (Figure
\ref{fig:isolatedsacs}C).

The dense synaptic connectivity of the SAC network and the potential role for extra-synaptic transmission
suggests that a model based on local coupling between cells is appropriate.
As such, coupling between amacrine cells is modeled via the
volume release and diffusion of acetylcholine (ACh), and an excitatory
post-synaptic current dependent on the local concentration of ACh.
By taking a continuum limit of this amacrine cell network, a reaction-diffusion
model is thus described (Methods). By modelling lateral excitation as a diffusion process the model is mathematically tractable. It should be emphasized that the diffusion of ACh is best thought of as an effective diffusion process, representing the combination of synaptic and extra-synaptic excitation through acetylcholine.
Below we will demonstrate the inclusion of long-range connections, in addition to local, `diffusive', ones has minimal impact in simulations.

The model is described in more detail in the Methods, but its notation
is established here. For a SAC located at $x$, its membrane potential
at time $t$ is described by $V(x,t)$. Three dynamic variables regulate the membrane potential:
$A(x,t)$, the extra-cellular ACh concentration which provides an excitatory
current to the SAC; $S(x,t)$, a calcium-dependent potassium channel which provides an inhibitory,
slow after hyperpolarization (sAHP) current; and $R(x,t)$, an inhibitory potassium channel modeled as
in the original Morris-Lecar model. The vector 
\begin{equation}
\mathbf{v}=(V,A,R,S)\label{eq:statevars}
\end{equation}
 then specifies the state of the system and obeys the following dynamics
\begin{align*}
\mathbf{v}_{t} & =\mathbf{f}(\mathbf{v})+\mathbf{D}\nabla^{2}\mathbf{v}.
\end{align*}
The matrix $\mathbf{D}$ is a diagonal matrix whose specified entries indicate
the diffusion coefficient of each dynamic variable
\[
\mathbf{D}=\left(\begin{array}{cccc}
0\\
 & D\\
 &  & 0\\
 &  &  & 0
\end{array}\right),
\]
and whose blank entries represent zeros. The function $\mathbf{f}(\mathbf{v})$ specifies the cell-intrinsic dynamics.

\begin{figure}[!ht]
\begin{center}
\includegraphics{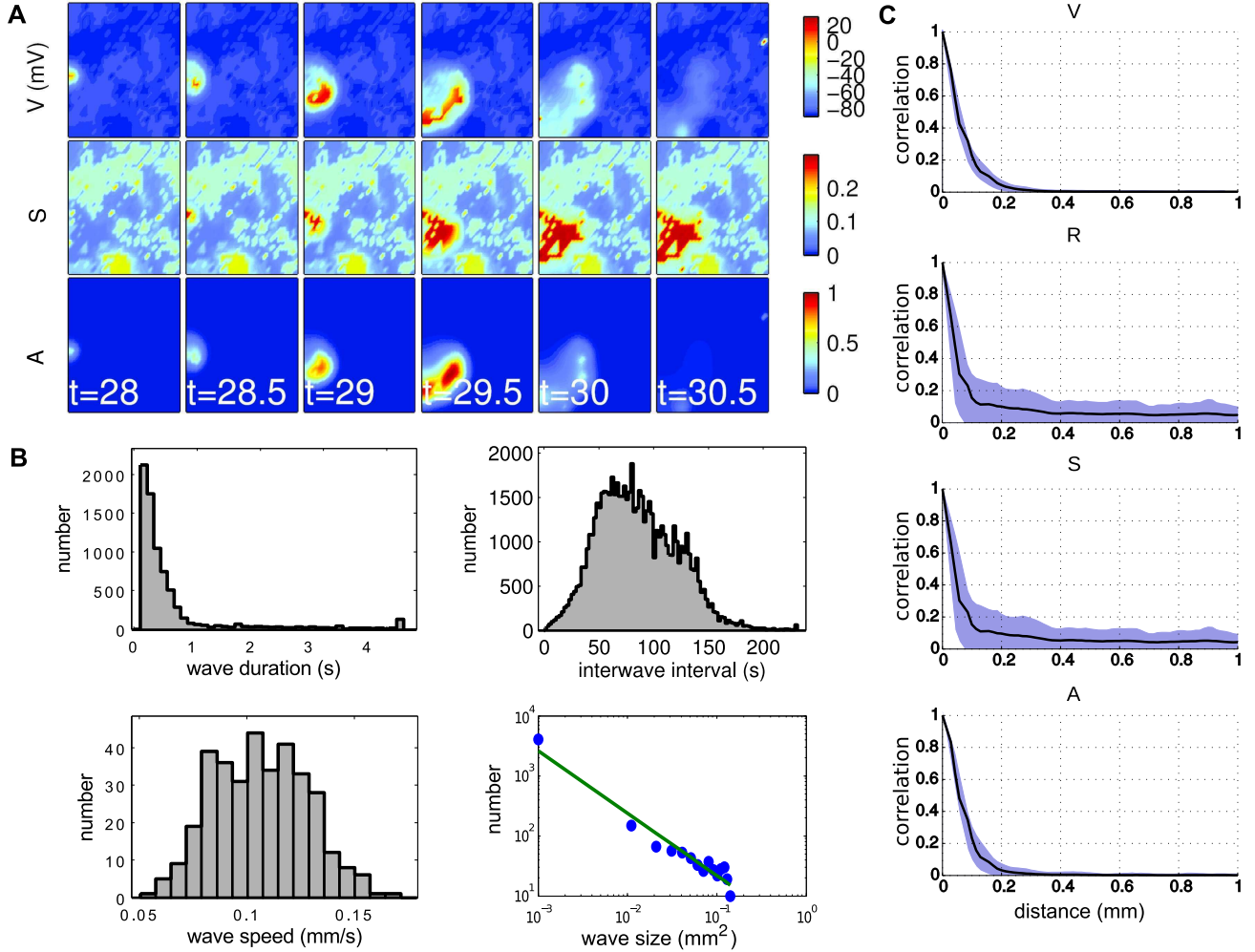}
\end{center}
\caption{\textbf{Model produces realistic cholinergic waves.}
\textbf{A.} Network dynamics showing spatiotemporal patterns of retinal
waves \textbf{B.} Distribution of wave sizes, speeds, durations and
inter-wave intervals from 2500s of simulation. Mean wave size is 0.017
mm$^{2}$($\pm$0.059mm$^{2}$), mean wave speed is 0.11mm/s ($\pm$0.022mm/s),
mean wave duration is 0.63s ($\pm$0.90s), and mean inter-wave interval
is 49s ($\pm$25s). \textbf{C. }SACs exhibit variable participation
in waves. Pearson correlation coefficient between a cell in the center of the domain and all other cells. The correlation coefficient for each variable is plotted as a function of euclidean distance between cells. Computed using one 2500s simulation, with activity recorded every 0.1s. Solid curve represents a loess moving average estimate of mean correlation as a function of distance. Shaded region highlights all points within one standard deviation of this mean curve.}
\label{fig:realwaves}
\end{figure}

Simulations show qualitatively that the model produces realistic waves
(Figure \ref{fig:realwaves}A; Movie S1). Waves propagate
without bias in their initiation region or direction (due to the translational
and rotational symmetry of the equations), occur on average once every
60s, propagate at an average speed of 150 $\mu$m per second, and
exhibit a broad distribution of wave sizes, all of which are consistent
with \emph{in vitro }recordings of mice retinal waves \cite{Ford2012}
(Figure \ref{fig:realwaves}B). The uniform distribution of initiation
points is expected given the homogeneity of the connectivities (diffusion
coefficient) of the model. Some studies show that both \emph{in vivo} and \emph{in vitro} recordings
contain a directional bias in propagation \cite{Ackman2012, Stafford2009}, which could be modelled with a drift-diffusion model. However, since there is presently no physiological model for how this directionality occurs, we do not attempt to address these issues here. 

The spontaneous firing of each cell is modeled as a Poisson process, which makes it simple
to fix the per-cell spontaneous activation rate to be consistent with
the recordings of Ford \emph{et al }2012 \cite{Ford2012}. The mean firing
rate is set to be low, such that adjacent spontaneous activations
are exceedingly rare, and hence waves are the result of a single SAC
depolarization. However, the mean rate is sufficiently high to desynchronize
the network, such that the correlation of activity of neighboring
SACs quickly decays as a function of distance (Figure \ref{fig:realwaves}C). Indeed, the correlation of both refractory variables ($S$ and $R$) decays faster than either the voltage or acetylcholine variables, indicating neighboring SACs can exist in different refractory states and thus exhibit variable participation in waves \cite{Ford2012,Hennig2009}.
These features are consistent with physiological waves observed in
mice. Further, by assuming a continuum model, the dimensionality of
the dynamics can be drastically reduced, when compared with a network model.
This feature is exploited in the following section.

\subsection*{Necessary conditions for wave propagation}
 
Having established the model produces realistic waves, in this section
we determine under what conditions propagating activity exists.
For this note that previous studies predict that the evolving mosaic of retinal wave activity is a result of the refractory period
of SACs, which delineates shifting boundaries of future wave activity
\cite{Ford2012}. If previous wave activity is indeed responsible
for the finite spatial extent of retinal waves then, were the amacrine
cell network in a homogeneously recovered state, activity would spread
across the entire retina without dissipation. Under this hypothesis,
a prerequisite for the amacrine cell network supporting propagating
retinal wave activity is thus that it should support traveling wave
solutions were it provided an infinite spatial domain at rest -- or
that it should be an excitable medium. 

\begin{figure}[!ht]
\begin{center}
\includegraphics{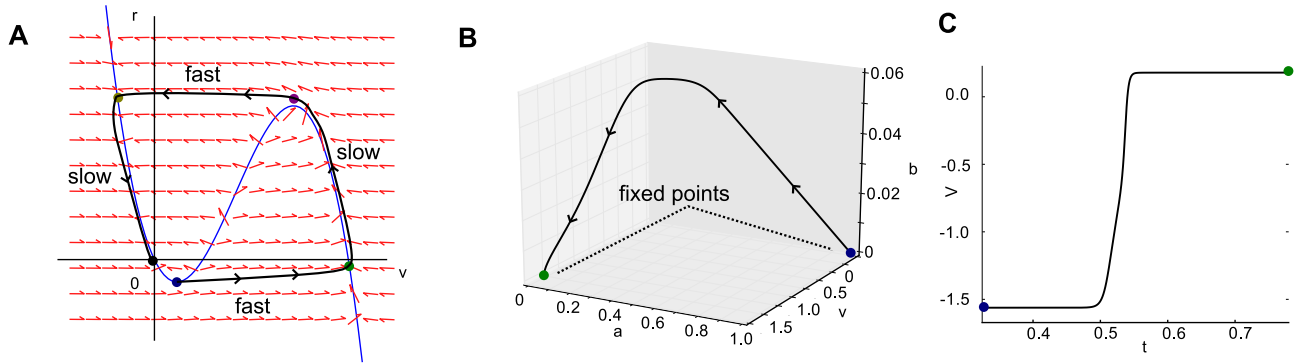}
\end{center}
\caption{\textbf{Construction of traveling wave-front.}
\textbf{A.} Fast-slow dynamics in the canonical Fitzhugh-Nagumo model
of action potential generation. Black curve represents a trajectory of an action potential through phase space, in which a fast transition occurs between the rest
(blue dot) and excited state (green dot), followed by slow excited dynamics (green to purple dot), another fast transition
between the excited and refractory state (purple to yellow), and slow dynamics while refractory (yellow to black). Red arrows represent flow lines, and the blue curve is the $V=0$ nullcline which defines the slow manifold ($R=0$ nullcline not drawn for clarity). \textbf{B. }The fast system
here is described by three dynamical variables ($v$, $a$, and $b\equiv a'$).
Shown here is the trajectory connecting the rest (blue) and excited
(green) fixed points, defining the wavefront. \textbf{C.} Temporal
voltage dynamics of the wave front.}
\label{fig:travellingfront}
\end{figure}

To investigate parameters in which our retinal waves model is excitable
is this sense, the amacrine cell network is studied as a non-linear
reaction-diffusion system, in which we seek parameters under which
traveling wave solutions can be constructed. The method is outlined
briefly here, and described in further detail in Methods. We take
advantage of the fact that the voltage and acetylcholine variables
change on a faster timescale than either refractory variables. Following
a singular-perturbation analysis outlined by Keener and Sneyd, 2001 \cite{Sneyd2001},
(treated in a number of other texts also, \emph{e.g.} Ermentrout and
Terman 2010 \cite{Ermentrout2010a}), the dynamics can thus be broken
into a fast and slow system:

\textbf{
\begin{align}
v_{t} & =f(v,r,a),\nonumber \\
r_{t} & =\epsilon h(v,r,s),\nonumber \\
s_{t} & =\epsilon\theta k(s,v),\nonumber \\
a_{t} & =\nabla^{2}a+\frac{k(a,v)}{\tilde{\tau}_{ACh}},\label{eq:dimlesssystem-1}
\end{align}
}where the lower-case variables denote dimensionless quantities corresponding
to their upper-case equivalents of Equation \ref{eq:statevars}. Singular perturbation analysis is used to study systems for which dynamics on more than one timescale need to be considered. For such systems, attempting to find approximate solutions by neglecting very fast or very slow components would result in an incorrect description of the dynamics. Here, the
parameter $\epsilon$ represents the ratio of the fast and slow timescales
and is to be considered small (for the default parameters used in simulations $\epsilon \approx 0.001$). The model requires singular perturbation techniques because the small parameter affects the leading order derivative terms.  The fast-system models the dynamics
during the jump from the rest state to the excited (depolarized) state, or from
the excited state to the refractory state. In one spatial dimension
$\nabla^{2}=\partial_{xx}$, and the zeroth-order ($\epsilon\to0$)
dynamics are therefore
\begin{align*}
v_{t} & =f(v,r,a),\\
a_{t} & =a_{xx}+\frac{k(a,v)}{\tilde{\tau}_{ACh}}.
\end{align*}
In a moving frame with speed $c(r)$, stationary solutions which connect
the rest and excited fixed points are \emph{heteroclinic} orbits which
represent traveling front solutions. The basic idea of the wave front construction is outlined in
Figure \ref{fig:travellingfront}.

\begin{figure}[!ht]
\begin{center}
\includegraphics{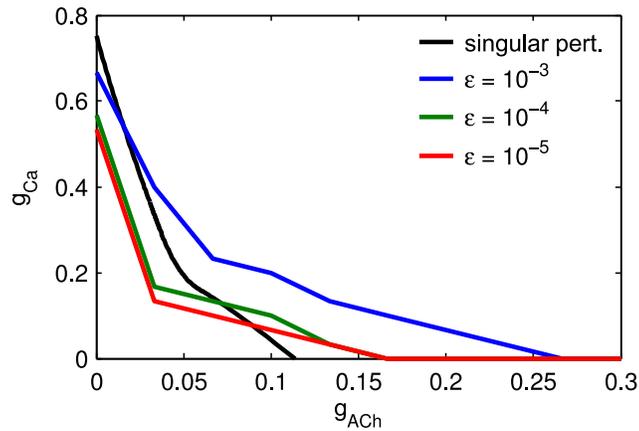}
\end{center}
\caption{\textbf{Parameter regimes which produce propagating activity.} Numerical determination
of retinal wave excitability threshold for different timescales $\epsilon$ and excitability
threshold determined through singular perturbation analysis,
both as functions of $g_{ACh}$, $g_{Ca}$. Each point on each
curve indicates a point in parameter space in which the wavefront
transitions from propagating to receding. Points in parameter space
below each curve are therefore not excitable, while those above are excitable.}
\label{fig:excitabilitythreshold} 
\end{figure}

Parameters for which a heteroclinic orbit exists for a positive speed
$c$ are parameters which support propagating activity. Parameters
at the transition between a propagating, positive speed, traveling
front and a receding, negative speed, traveling front (that is, when
$c=0$) represent an excitability threshold. Figure \ref{fig:excitabilitythreshold}
demonstrates this excitability threshold over a two dimensional
parameter space, computed using the above framework and, for comparison, computed using
numerical simulations of the model for different values of $\epsilon$ (the separation of fast and slow time scales). The analysis and simulations exhibit the same general behavior. The analysis method presented here is a more direct method for determining wave propagation because it is not necessary to set up and solve numerically
a solution for each point in parameter space to be queried. Refer to Methods for more detail on the numerical simulations.

Both analysis and simulation demonstrate that excitability is sensitive
to the value of both maximal conductances $g_{ACh}^{M}$ and $g_{Ca}^{M}$. Further, excitability
is independent of all aspects of the model not involved in the fast
dynamics. This means that neither the spontaneous activation
rate nor the duration of either refractory variable have an effect
on wave propagation. Indeed, this is consistent with data presented
in Ford \emph{et al} 2012 \cite{Ford2012}, which shows manipulations affecting
the strength of the sAHP current have minimal effect on measured wave
speed. In the sense defined here, excitability is also independent
of the diffusion coefficient $D$, as it is scaled out of the dimensionless
equations used to compute the excitability thresholds. Thus, beyond assuming a
non-zero coefficient, the diffusion rate merely determines the speed
at which waves propagate. The existence of spontaneous, propagating
activity in the developing retina is thus determined by factors regulating
individual SACs excitability and their synaptic strength.

It is important to understand that this analysis is only for the case of a 
domain uniformly at rest. Of course, we would like to consider the existence 
of waves within a medium that is not uniformly recovered but 
for which some of the domain may be refractory from previous activity. Our analysis provides a necessary condition for the existence of propagating 
activity in this more general case: a network incapable of supporting wave propagation 
at rest is a network incapable of supporting wave activity when some of its cells are refractory.
In this more general case, waves propagate not only if SACs are sufficiently excitable,
but also if neighboring SACs are sufficiently recovered from prior depolarizations. 
The wave speed is then modulated by the refractory state as well (\emph{e.g.} Figure \ref{fig:pharma}). 

\subsection*{Waves under physiological manipulations}

The pharmacological or genetic manipulation of retinal waves forms
a major component of their experimental study. In this section we show how two such common manipulations are recapitulated by our model and
use the above analysis to interpret the effects of these manipulations. Using a biophysically inspired model is beneficial 
since model parameters have clear, experimentally determinable values. 

\subsubsection*{Synaptic connection strength}

We first investigate the effect of nicotinic acetylcholine receptor (nAChR)
agonists/antagonists by varying the maximal synaptic conductance $g_{ACh}^{M}$ (see Methods for definition).
Our analysis shows both the wave speed at rest and the wave speed
as a function of refractory state $R$ are highly sensitive to changes
in $g_{ACh}^{M}$ (Figure \ref{fig:pharma}A). A 25\% reduction in
$g_{ACh}^{M}$ lowers the $C(R)$ curve significantly, making the
medium less excitable and waves more easily blocked by encounters
with refractory cells. This manifests in simulations as a reduction
in wave size, and a change in the inter-wave interval: from sharply
peaked at a mean of approximately 50s to a monotonically decaying
function similar to the individual SAC spontaneous firing rate (Figure
\ref{fig:isolatedsacs}), indicating that wave activity is localized
and activations are primarily caused by spontaneous activity, not
wave activity. This is indeed observed in pharmacological studies of
mice, chick and turtle retina \cite{Sernagor2000,Sernagor2003,Bansal2000},
which produce both a decrease in wave frequency and a decrease in
wave size following treatment with nAChR antagonists. 

Additionally, a study by Xu \emph{et al} 2011 \cite{Xu2011} used transgenic mice in 
which only about half of SACs express functional nicotinic acetylcholine 
receptors, which reduces the effective coupling of the network. In
these mice, wave frequency and speed were unchanged, whereas wave size was 
significantly reduced, as the analysis of our model predicts.

Conversely, a 25\% increase in $g_{ACh}^{M}$results in an increase
in wave speed both from rest and as a function of refractory state,
indicating that wave activity is more robust and less likely to be
blocked by encounters with refractory cells. Indeed, in simulations
this results in an increase in wave size and a more frequent and sharply
peaked inter-wave interval distribution. 


\subsubsection*{Modulation of sAHP current}

\begin{figure}[!ht]
\begin{center}
\includegraphics{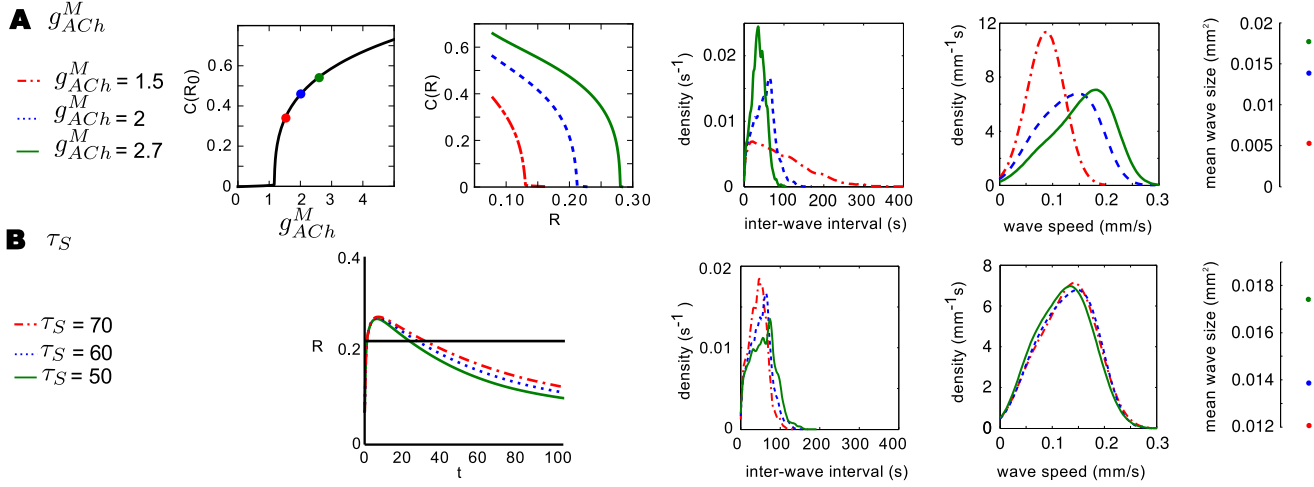}
\end{center}
\caption{\textbf{Modeling biophysical manipulations.} \textbf{A
.}Synaptic connection strength $g_{ACh}^{M}$ is varied. Sub plots
from left to right: speed of wave front $c(R_{0})$ at rest (when
$R=R_{0}$) as a function of conductance $g_{ACh}^{M}$, velocity
indicates maximum wave-front speed since $R\ge R_{0}$ and $C(R)$
is monotonically decreasing, point at which $c$ becomes zero represents
excitability threshold; wave-front speed as function of refractory
variable $R$ for three different values of $g_{ACh}^{M}$; from 5000s
of simulation of model with indicated values of $g_{ACh}^{M}$ interwave-inteval;
wave speed distribution; and mean wave size. \textbf{B.} Sub plots
from left to right: dynamics of refractory variable $R$ of individual
SAC following depolarization with different sAHP timescales $\tau_{S}$,
black line indicates refractory value above which $C(R)\le0$ and
thus represents an absolute refractory time period in which SAC is
not sufficiently excitable to participate in future wave activity;
from 5000s of simulation of model with indicated values of $\tau_{S}$inter-wave
interval; wave speed distribution; and mean wave size.}
\label{fig:pharma}
\end{figure}

We also investigate the effect on waves from varying the sAHP current. Experimentally, this is achieved through stimulating the cAMP second-messenger pathway via forskolin. We study this effect by varying the sAHP timescale $\tau_{S}$. Since $\tau_{S}$ does not effect the fast time scale dynamics, our singular perturbation analysis does \emph{not} provide insight into how $\tau_{S}$ effects wave speed or propagation. However, using the model's excitability criteria does allow us to compute the duration of an absolute refractory period, which is the length of the period in which activation, either spontaneous or through incident wave activity, results in subsequent a refractory state that does not support a propagating wave. This period is shown to be on the order of 30 seconds, in accordance with experimental findings \cite{Bansal2000}. (Figure \ref{fig:pharma}B)

Retinal wave simulation \emph{does} show how $\tau_{S}$ affects
wave properties. An increase (respectively decrease) in $\tau_{S}$ results
in an increase (respectively decrease) in the interwave interval and a minimal
change in wave size and wave speed (Figure \ref{fig:pharma}B). The
reduction in IWI is expected given the decrease in absolute refractory
period. The minimal change in wavespeed is also expected. Further,
the moderate changes in $\tau_{S}$ presented here should not affect
significantly the wave size (larger changes in $\tau_{S}$ through
which different wave behavior regimes are explored are investigated
below). These results agree with the study of Ford
\emph{et al }2012 \cite{Ford2012}, which found that treatment with
1$\mu$M of forskolin resulted in roughly a 50\% reduction in IWI,
and only a moderate reduction in wave speed. The same increase in
wave frequency and decrease in underlying sAHP current are also observed
in ferrets and rabbit \cite{Zheng2006,Stellwagen1999}. 


Recently, Ford \emph{et al} 2013 \cite{Ford2013} proposed that the sAHP current is established by the two-pore
potassium channel TREK1. In this study TREK1 knockout mice are shown
to exhibit retinal waves with a significantly reduced (approximately
halved) interwave intervals. The present model, in which wave frequency is shown to be increased by a reduction of the timescale of sAHP recovery, is consistent with these findings. 

It is possible the effect of forskolin is on the strength of the sAHP current, not on its decay rate, thus we also investigated the effect of changing the strength parameter $\alpha$, which determines how sensitive each SAC is to the slow refractory state $S$. The effect was comparable to varying $\tau_S$ -- a decrease in wave frequency and slight change in wave speed and size following an increase in $\alpha$ (Figure S3).

In summary, the preceding sections have demonstrated that our model produces physiological retinal waves, and that the existence of propagating activity can be understood by studying the model as an excitable medium. This allows for the effect of common genetic and pharmacological manipulations to be predicted. 

\subsection*{Spatio-temporal patterns of propagating activity}

We now turn our attention to a second question: given parameters in which propagating activity exists, what model parameters determine their form? Indeed, the spatiotemporal patterns of cholinergic retinal waves are often
similar, despite differences in a range of physiological parameters,
in a range of species. How is this similarity of form maintained given
variations in physiology? Given parameters in which propagating activity
exists, this section addresses the nature
of the resulting spatiotemporal patterns. Hennig \emph{et al} 2009
\cite{Hennig2009} notes that, both in MEA recordings and computer
simulation, physiological waves take the form of avalanches, or power-law
sized events. Thus it has been proposed that the developing amacrine cell
network is in a critically configured state -- a transition between
a locally and globally connected network. Such neuronal avalanches
have been observed and extensively studied in the both \emph{in vitro
}\cite{Petermann2009} and \emph{in vivo }\cite{Beggs2003} cortex.

Here, we make an analogy between retinal waves and a model of self-organized
criticality (SOC) \cite{Bak1988}, allowing us to determine for which
parameters avalanches may be observed. A well-studied example of a
complex system potentially demonstrating self-organized criticality
is the Drossl-Schwabl forest fire model (DS-FFM) \cite{Drossel1992,Bak1990},
imagined as a grid in which each unit is either occupied (by a tree),
ignited (burning tree), or empty (ash, burnt tree). At each discrete
time step: 1) an occupied grid point spontaneously ignites with probability
$f$, 2) burning trees ignite their occupied neighbors, 3) burning
trees become empty and 4) empty sites regrow a tree with probability
$p$. The analogy between the forest and the retina is clear: amacrine
cells spontaneously fire at some rate and excite their nearest recruitable
neighbors to also fire, after which these cells experience a slow
recovery time, which proceeds at some average rate. 

For a simulated lattice of $n^{2}$ cells, representing $L^{2}$mm$^{2}$
of retina (in our simulations $n=64$ and $L=2\text{mm}$), a simple heuristic derivation shows a relationship between
$f$ and $p$ of the DS-FFM and the retinal wave model parameters:
\[
f=\frac{\pi n^{2}c^{2}\tau^{2}\hat{f}}{L^{2}},\quad p=\frac{\tau}{\rho},
\]
for wave speed $c$, a per cell spontaneous firing rate $\hat{f}$,
burst duration $\tau$, an effective refractory period $\rho$, and
length of domain $L$. These values are determined either directly
from model parameters ($\hat{f}$, $L$, $n$), from the model analysis ($c$ and $\rho$), or from simulation ($\tau$). 

The heuristic derivation is as follows: to estimate $f$ and $p$ from retinal
wave simulations, and relate it to the DS-FFM, we rescale time and space
so that one time unit is the duration of a burst during wave activity
($\tau$); in this way after one time unit every active grid point
is now refractory. And rescale space such that the area of one `lattice
point' is the area covered by all grid points excited as a result of one point
spontaneously firing (initiating a wave) ($\pi c^{2}\tau^{2}$); in
this way, during one time unit, an active lattice point induces the nearest (and only
the nearest) excitable lattice points to become active.
When scaled in this way, the dynamics of our retinal wave simulations
approximate the dynamics described by the rules of the DS-FFM. The
probability of 'regrowth' (recovery) is $p=\tau/\rho$, and the
probability of a lattice point spontaneously firing is assumed to be the probability
of a single simulated cell firing, multiplied by the number of
cells that are included in that lattice point ($\pi c^{2}\tau^{2}n^{2}/L^{2}$).

\begin{figure}[!ht]
\begin{center}
\includegraphics{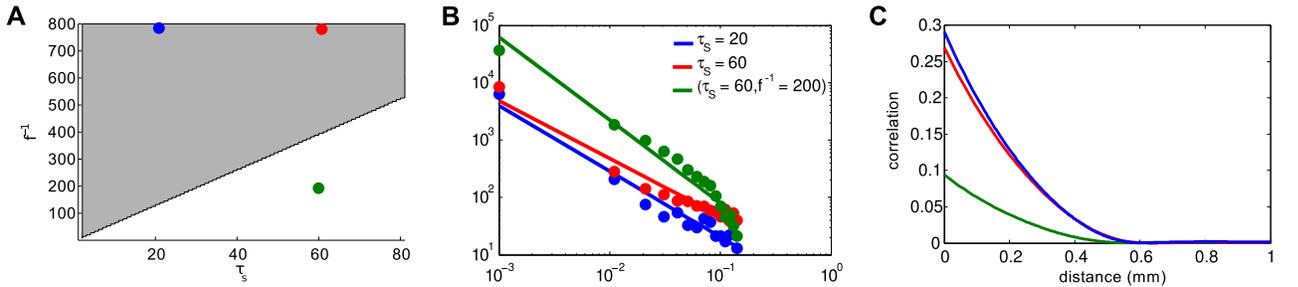}
\end{center}
\caption{\textbf{Power-law distributed wave-size retinal waves.} \textbf{A.} Parameter space in which avalanches are
expected (gray, Equation \ref{eq:soc_crit}) and three sample points\textbf{
B.} Wave size distributions (points) following 5000s of simulation
on a 128$\times$128 domain for specified values of $\tau_{S}$ and
$f^{-1}$. Solid lines represent log-linear least-squares lines of
best fit, having slopes: $\theta=-1.45$ ($R^{2}=0.95$, green), $\theta=-1.01$
($R^{2}=0.95$, red) and $\theta=-1.14$ ($R^{2}=0.95$, blue) \textbf{C.}
Correlation in membrane potential between cells of a given distance
apart.}
\label{fig:soc}
\end{figure}

In the DS-FFM, SOC is expected when \cite{Drossel1992}
\[
(f/p)^{-1/2}\ll p^{-1}\ll f^{-1},
\]
or, for the present model, when
\begin{equation}
\left[\frac{\pi n^{2}c^{2}\tau\hat{f}\rho}{L^{2}}\right]^{-1/2}\ll\frac{\rho}{\tau}\ll\frac{L^{2}}{\pi n^{2}c^{2}\tau^{2}\hat{f}}.\label{eq:soc_crit}
\end{equation}
In this regime, retinal wave, or forest fire, sizes are characterized
by a power-law distribution with scaling exponent of approximately
$\theta=-1.15$ (simulation based \cite{Grassberger2002}, theoretical
based \cite{Loreto1995,Hergarten2011}). Parameters which have the
largest and most direct impact on $f$ and $p$ are the per cell spontaneous
firing rate $\hat{f}$, and the slow refractory variable $\tau_{S}$
(refer to Figure \ref{fig:pharma}): these are the parameters which
best determine when criticality may be observed. Figure \ref{fig:soc}
demonstrates that, within the region described by Equation (\ref{eq:soc_crit}),
wave sizes distributions approximately follow a power law with an
estimated exponent close to the expected $\theta=-1.15$. Conversely,
simulations performed outside this parameter region are sub-critical
and do not follow an approximate power-law. The same behaviour is observed when the distribution of wave duration (or lifetimes) are considered (Figure S4). This is further made clear by
looking at the correlation in voltage activity between cells of a
given distance from one another (Figure \ref{fig:soc}C). Parameters
for which power-laws are observed produce an initially high, but sharply
decaying correlation function, while the sub-critical parameter set
produces significantly less correlated activity. This is indicative
of the smaller, more localized wave activity expected in a sub-critical
system. 

The relationship between the rate of spontaneous firing $\hat{f}$, and the slow
refractory variable $\tau_S$, dictates whether the network exists in a 
critical state. This simple inverse relationship (Figure \ref{fig:soc})
can explain how robust waves with similar spatiotemporal characteristics
exist in very different parameter regimes. Experiments in mice indicate
that SACs depolarize infrequently but exhibit sAHPs lasting as long as two
minutes \cite{Ford2012}, whereas experiments in rabbit \cite{Zheng2006}
show a spontaneous depolarization rate an order of mangitude higher and 
sAHPs that are substantially shorter. While these two systems have cellular
parameters that differ on an order of magnitude, their collective network
activity is very similar. This can be explained by the fact that the ratio
of $\hat{f}$ to $\tau_S$ is conserved across species.

This analysis can thus be used to predict, given particular parameters
for a particular species, whether or not the wave-size distribution
will be critical or sub-critical. Further, since
both parameters varied here, $\hat{f}$ and $\tau_{S}$, have no bearing
on the excitability of the medium, the analysis of the spatio-temporal
patterns of the amacrine cell network separates neatly into two stages.
Firstly, if the medium is not excitable then only small, localized activity is expected. If the medium is
excitable then, secondly, the spatio-temporal patterns expected will
depend on the value of both $\hat{f}$ and $\tau_{S}$, as discussed
here.

\section*{Discussion}

\subsection*{Comparison to previous studies}

Close collaboration with experimentalists means computational modeling
of retinal waves has been notably successful (refer to the reviews
Gjorgjieva and Eglen 2011 \cite{Gjorgjieva2011} and Godfrey and Elgen 2009 \cite{Godfrey2009}).
Models have helped identify SACs as the
cell layer which both initiates and laterally propagates waves \cite{Feller1997},
and have highlighted the importance of an activity-dependent refractory
period to the generation of physiological waves \cite{Godfrey2007}.
Most recently, the models of Hennig \emph{et al} 2009 \cite{Hennig2009}
and Ford \emph{et al }2012 \cite{Ford2012} (henceforth referred to
by their first authors) have investigated how the cell-intrinsic spontaneous
firing rate and the duration of each cell's refractory period contributes
to physiological waves.

Godfrey and Hennig both propose the importance of a refractory period that scales with the degree of excitation. The Godfrey model was deterministic, and thus was
reliant on the activity dependent refractory periods to make the dynamics
chaotic and therefore consistent with observed waves -- variable in
size, and with shifting wave boundaries. 

In the Hennig model, the competition between the synchronizing force
of waves and the desynchronizing force of spontaneous activity is balanced
at physiological wave parameters. This was adjusted in the Ford model
to match the observed low spontaneous activation rate observed in
mice SACs. Physiological waves were then only observed when each cell's
refractory time scale was allowed to vary randomly (a hypothesis also
pursued in Feller \emph{et al }1997). Ford hypothesizes that this
cell to cell variability is thus a necessary component of the developing
amacrine cell network. 

We find that neither hypothesis is
necessary in producing realistic waves -- our model does not contain
an activity-dependent refractory variable or cell to cell variability in parameters. Our model is stochastic, consisting of sparse spontaneous depolarizations with enough strength to initiate a wave by themselves. These strong, sparse depolarizations are sufficient to induce variability in refractory periods. It remains to be shown how this might change in other species whose spontaneous depolarization rate is much higher, and in which cell-cell variability or activity-dependent refractory periods may play a more important role.

\subsection*{Experimental validation of the model}

The model makes several predictions. Firstly, it provides a way to systematically investigate the effect of different parameters on both wave propagation and spatiotemporal patterns. One way to confirm the model experimentally would then be to match (more closely than was able to be performed in the present study) the parameters of the model to observed wave statistics such that in a quantitative fashion the effect of particular pharmacological manipulations (\emph{e.g.} nAChR antagonists) could be reliably reproduced. The test would then be to predict the outcome of a different pharmacological manipulation (\emph{e.g.} cholinesterase inhibition) that the model was not calibrated against. 

A simpler validation would be the following: a prediction the model makes is that certain parameters affect only the existence of propagating activity, while certain parameters affect only the form of that propagating activity (power-law distributed or otherwise). Thus manipulations that affect only parameters within one of these classes should have no effect on behaviours determined by parameters in the other class. For instance, the only effect changes to parameters of the sAHP current, according to our analysis, is on the form of the activity -- spiral waves, large domain-covering waves, \emph{etc} -- not on their existence. 

Finally, the model predicts the existence of power-law distributed activity within a particular parameter range based on an analogy to a forest fire model. There are many different models of self-organized critical systems which have slightly different rules and slightly different statistics, and it is not clear that the forest fire model indeed is the best analog. These different models exhibit different sub-sampling artifacts, when only a subset of cells are observed, and these effects thus provide a way of teasing apart different SOC candidates \cite{Priesemann2009a}. Examining sub-sampling effects in our model's simulations, and in high-density recordings of \emph{in-vivo} or \emph{in-vitro} retinal waves, would provide a way of determining which model exhibiting criticality is most appropriate.

\subsection*{On the extra-synaptic diffusion of ACh}

Previous retinal wave studies have hypothesized an extra-synaptic
agent as responsible for wave propagation: Burgi \emph{et al }1994
\cite{Burgi1994}, for instance, noted that the relatively slow wave
speed is consistent more with an extra-synaptic agent than either
gap junction or direct synaptic stimulation, and proposed extracellular
potassium as a source of lateral excitation. This particular hypothesis
was later discredited. However diffuse action of ACh within the IPL
has been observed coincident with wave activity, suggesting that the
volume transmission of acetylcholine may instead play a role \cite{Ford2012,Syed2004}.
Thus, although direct recurrent synaptic connections between SACs
both in mice and in rabbit have been observed \cite{Zheng2004,Ford2012},
there is evidence that the diffuse release of ACh is responsible for
wave propagation: synaptic currents recorded during waves last several seconds
beyond depolarization, suggesting excitation via an extra-synaptic
agent; and retinal waves drive ON and OFF RGCs \cite{Wong1996} and
cells in the inner nuclear layer \cite{Wong1995}, despite the restricted stratification of SAC processes within the IPL that do not form direct contacts with all cells, again suggesting
propagation via volume transmission.

It is worth pointing out that extra-synaptic neurotransmission is
observed throughout the developing nervous system to produce correlated
spontaneous activity, and thus may be a general phenomenon regulating
a diverse range of developmental programs. This may happen when the
development of neurotransmitter release through synaptic vesicle fusion
precedes the expression of the uptake transporters and breakdown enzymes
required to prevent spillover into the extra-cellular space. For a
the review of the role of extra-synaptic stimulation in generating
spontaneous activity refer to Kerschensteiner 2013 \cite{Kerschensteiner2013}.

The role of volume transmission has been explored in cortex \cite{Allene2008},
hippocampus \cite{Cattani2007a}, as well as extensively in stage III
(glutamatergic) retinal waves \cite{Akrouh2013a,Blankenship2009,Firl2013}.
In stage III waves extra-synaptic glutamate mediates lateral connections
between bipolar cells.

The model presented here represents a theoretical validation that
volume release of ACh is able to mediate stage II retinal waves. However,
further experimental validation is needed beyond the above observations
and this model. It should therefore be emphasized that the motivation
of writing a model based on diffusion of acetylcholine is not to test
the role of volume transmission. The dense lateral connectivity observed
between SACs (synaptic or otherwise) means that lateral excitation
is well approximated by a diffusion process -- the model's diffusion
of acetylcholine should be interpreted as a phenomenological, or effective,
diffusion, not as a literal diffusion process based on the net Brownian
motion of molecules which allows for singular-perturbation
analysis to be performed. 

Indeed, the extent of a SAC's arborization is only approximately 100 microns, so that even if synaptic terms were explicitly included their connectivity would be well-modelled by a short-range Gaussian weight function similar to the Gaussian kernel that results from modelling interactions as occurring through a diffusion process \cite{Whitney2008}. The only cells in the retina with long range connections are other types of GABAergic amacrine cells, which do not alter waves since GABA antagonists play a minor role in shaping the spatiotemporal features of waves. For these reasons models with long-range connectivity were not pursued in this study.

\subsection*{Stage II retinal waves and development}

Retinal waves are one example of spontaneous activity in the nervous
system instructing developmental processes. A number of processes
appear to rely on the activity generated by retinal waves. Indeed, they
have been shown to coordinate activity throughout the visual system \cite{Ackman2012}. In animals in which waves are blocked,
the formation of the retinotopic map \cite{Chandrasekaran2007}, the
separation of retinogeniculate projections into eye-specific layers
of the LGN \cite{Xu2011}, and the formation of ocular dominance and
orientation selectivity \cite{Huberman2008} (and references therein)
are all adversely effected. 

Retinal waves are thus believed to provide a naturalistic stimulus
from which downstream connections can be refined and stablized. However,
it remains unclear how important the spatiotemporal patterns of the
activity are, or whether any activity, regardless of its properties
(size, speed, typical shape), would suffice to provide input for correct
development to occur. Hennig \emph{et al} 2009 \cite{Hennig2009}, for instance, suggests
that the critical form of the retinal waves are a useful feature,
because, by definition, critical-state activity consists of events
possessing no intrinsic length scale, they thereby provide an input
without bias to any particular feature length. Godfrey \emph{et al} 2009 \cite{Godfrey2009a},
on the other hand, investigates this question explicitly in a model
of retinotopic map refinement and finds that the only feature which
affects the degree of refinement is the spatial correlation activity
function -- the size, speed and frequency of the waves otherwise had
no effect.

If correlation structure of retinal activity is the main factor regulating
downstream developmental processes, then how is this structure dependent
on the spatiotemporal properties of retinal waves? Are scale-free,
critical `avalanches' the only form of activity that provide the necessary
correlation structure? In other areas (\emph{i.e.} cortex), a network
operating at a critical state can be shown to be optimal for information
processing, and for maintaining a high dynamic range \cite{Beggs2008,Shew2009}:
there is some functional role for criticality (as discussed in Shew
and Plenz 2013 \cite{Shew2013}), but its significance remains controversial
\cite{Beggs2012}. Retinal waves provide a useful study into the functional
role of criticality, as they occur within well defined and understood
circuitry where the above questions can be asked and thus critical
activity can (potentially) be assigned a specific developmental role.

\subsection*{Self-organized criticality}

Our study focused in particular on a forest fire model extensively
studied in the context of complex systems and critical phenomena for
its potential demonstration of self-organized criticality (SOC). This
connection between correlated activity within the central nervous
system and the DS-FFM has been noted before (\emph{e.g.} Buice and
Cowan 2009 \cite{Buice2009}). The DS-FFM model is not without its
issues: while initial theoretical results \cite{Drossel1992,Loreto1995}
provided justification for numerically observed scaling behavior,
subsequent, more extensive, numerical analysis \cite{Grassberger2002}
demonstrate the power-law scaling behavior to be transient and not
reflective of the `true' asymptotic behavior. Indeed, a variety of
scaling parameter ($\tau$) estimates have been observed for different
parameter values and lattice sizes (refer to the summary presented
in Pruessner 2012 \cite{Pruessner2012}).

In theory, critical behavior is expected as the ratio $\theta=f/p$
tends to infinity. However, this limit corresponds to lightning strikes
becoming increasingly rare, such that in simulations, for sufficiently
large $\theta$, strikes are rare enough that the entire domain recovers
by the time the next strike hits, and the resulting dynamics are simply
that the fire spreads across the entire domain. Thus, in practice,
critical behavior is to be expected when $\theta$ is large, yet the
characteristic length scale remains small compared to the lattice
size. In addition to this behavior, for sufficiently large $p$ values
spiral wave fronts are observed \cite{Clar1996}. These behaviours
-- propagating `critical', stationary rotating spirals and radially
symmetric domain-covering waves are the three regimes of spatially
extending activity observed in simulation of the present model, and
as classified in previous models \cite{Butts1999}. The latter two
behaviours are likely to occur in parameter regimes where the spontaneous
activation rate is sparse, thus creating large scale structures, and
where, relative to the wave-front speed, the rate of recovery
is either too fast (spiral waves) or too slow (domain-covering radial
waves). Though more systematic study is needed in these cases, we
argue the principle value of the analogy to the DS-FFM is not evidence
of critical behavior in the developing retina (as it is not clear the
DS-FFM is critical in a strict sense), but is the insight it provides
into when and how these different regimes will be observed.

\subsection*{Wider context}

This work shows that a model based on lateral excitation through the
diffusion of ACh can account for physiological retinal waves, and
can predict how wave dynamics consequently depend on biophysical parameters.
Specifically, by determining when the retinal network is excitable
and through analogy to a canonical forest-fire model, the spatiotemporal
patterns observed in different parameter regimes can be understood.
This framework is shown to be consistent with previous experimental studies.

We note that the nature of this analysis is novel: the singular perturbation
and traveling front construction analysis presented here has not been
extended before, from the well studied and more stereotyped waves studied
in cortex, to the considerably more complicated wave behavior observed
in the developing retina. 

A number of avenues for future work present themselves. First, the reaction-diffusion component of
the model can be naturally applied to glutamatergic waves, for which
there is more significant evidence that diffusion is indeed a key
excitatory process. Mechanistic models of stage III waves have recently
been developed \cite{Akrouh2013a,Firl2013}, and would benefit from
computational modeling to establish their validity. It is proposed in stage III waves that lateral excitation occurs via the diffusion of extra-synaptic glutamate released by bipolar cells in the ganglion cell layer, and that GABAergic connections from amacrine cells to bipolar and ganglion cells provides a source of inhibition that is absent during stage II waves. Changes to the present model to a model of stage III waves would therefore be significant, however the framework to both analyze and simulate the resulting model would remain the same. 

Second, the connection between between criticality and development can also be explored in much more depth than was performed in this study.

We believe the approaches developed here may also be applied to other forms of spontaneous and correlated activity in the nervous system.

\section*{Models}

\subsection*{The model }

SACs are assumed to obey dynamics based on a Morris-Lecar \cite{Hennig2009} model, which includes a quasi-stationary voltage dependent calcium conductance. The
voltage dynamics are described by 
\begin{align*}
C_{m}V_{t} & =-g_{Ca}(V-V_{Ca})-g_{K}(V-V_{K})-g_{L}^{M}(V-V_{L})-g_{ACh}(V-V_{syn})-g_{n}^MN(V-V_{Ca}),
\end{align*}
where 
\begin{align*}
g_{Ca}(V) & =\frac{1}{2}g_{Ca}^{M}\left[1+\tanh\left(\frac{V-V_{1}}{V_{2}}\right)\right],\\
g_{K}(R) & =g_{K}^{M}R,\\
\Lambda(V) & =\cosh\left(\frac{V-V_{3}}{2V_{4}}\right),\\
R_{\infty}(V) & =\frac{1}{2}\left[1+\tanh\left(\frac{V-V_{3}}{V_{4}}\right)\right],
\end{align*}
are the standard Morris-Lecar auxiliary functions. The model also includes an ACh conductance ($g_{ACh}(A)$) which depends
on the local concentration of acetylcholine ($A$): 
\[
g_{ACh}(A)=g_{ACh}^{M}\frac{\delta A^{2}}{1+\delta A^{2}}.
\]
Additionally, a slow AHP variable ($S$) is activated according to a voltage-dependent
function $G(V)$ with time scale $\tau_{ACh}$: 
\begin{align*}
S_{t} & =\gamma G(V)-\frac{S}{\tau_{S}}\\
G(V) & =\frac{1}{1+\exp[-\kappa(V-V_{0})]}.
\end{align*}
The sAHP variable enters the dynamics through its effect on refractory variable $R$ as follows 
\[
\tau_{R}R_{t}=\Lambda(V)(R_{\infty}-R)+\alpha S(1-R), 
\]
where $\Lambda(V)$ is defined above as part of the Morris-Lecar equations.

Cells are indexed by a continuous spatial parameter $x$ such that the
state space is described by the tuple $(V(x,t),R(x,t),S(x,t),A(x,t))$,
where $V(x,t)$ is the membrane potential. On spiking, cells release
ACh at a voltage dependent rate $\beta G(V)$ which diffuses with
coefficient $D$: 
\begin{align*}
A_{t} & =D\nabla^{2}A+\beta G(V)-\frac{A}{\tau_{ACh}}\\
G(V) & =\frac{1}{1+\exp[-\kappa(V-V_{0})]}.
\end{align*}
Coupling between SACs occurs only through the diffusion of ACh. 

To induce spontaneous
depolarizations, a noisy, excitatory current is included in simulations:
\[
g_{n}^{M}N(V-V_{Ca}),
\]
where at each time-step $N$ is a Bernoulli random variable ($N\sim B(1,p)$).
Refer to Table \ref{tab:parameters} for the parameter values used.
The value of $p$ is chosen such that the mean waiting time between sponataneous
noisy channel openings matches approximately the rate of intrinsic spontaneous bursts observed in mouse retina
recordings \cite{Ford2012}. 

\subsection*{Computational methods}

Numerical integration was performed using a time-splitting method,
in which the diffusion term was computed using a locally one dimensional
Crank-Nicolson method and the reaction term was computed using a two-stage
Runge-Kutta method. Simulations were computed in MATLAB (version 7.14.0;
R2012a. Natick, Massachusetts: The Mathworks, Inc., 2012) using a
64$\times$64 grid with a fixed step size of 1ms. The stochastic conductance
variable was updated every 10ms. Simulation data used in statistical
analysis was from 2500s of simulation on a 4mm$^{2}$ domain ($L=2\text{mm}$), following
a 500s warm-up period. The dimensions chosen mean that one grid point
has a length of approximately 30$\mu$m -- roughly the density of
SACs in the inner plexiform layer \cite{Butts1999}. Simulations were
performed on a dual 6-core Intel Xeon 3.07Ghz machine with 24GB RAM,
running Ubuntu 12.04. MATLAB code to run simulations is available for download at: https://github.com/benlansdell/retinalwaves.

\subsection*{Simulation statistics}

Statistical analysis is performed as follows: snapshots of the simulation
are taken every 10 time steps (10ms). To remove boundary effects,
grid points 5 units or less from the boundary are not considered for
analysis.  All grid points whose potential is above a threshold of
-60mV are labeled as active. All active points adjacent to one another
are assigned a common wave number. For each set of cells assigned
that number: the time from the first active cell to the last active
cell is its wave duration, and the total number of cells is its size.
The wave speed is calculated as in Blankenship \emph{et al} 2009 \cite{Blankenship2009}:
the wave initiation and termination points are identified and the
path the wave front travels between these two points is computed.
This allows the maximum wave front speed for that wave to be computed,
and the distance this wave front travels divided by the wave duration
gives an average wave speed. To give better estimates of the wave
front speed, waves which involve the activation of fewer than 50 cells
or whose duration is less than 1s are ignored, and waves which do
not have a distinct initiation point and are therefore the result
of two waves colliding are also omitted (Figure S1).
The interwave-intervals are computed by considering all above threshold
wave activity, collisions or otherwise, and, for each grid point,
measuring the time between successive threshold crossings. A minimum
IWI of 2s was imposed. 

\subsection*{A note on thresholds and smoothing voltages}

Previous models of retinal waves consisted of two layers: an amacrine
cell layer, which initiated and laterally propagated activity, and
a ganglion cell layer, which acted as a smoothed 'read out' layer
\cite{Feller1997}. Subsequent models did away with the ganglion cell
layer, and modeled only SACs \cite{Godfrey2007,Hennig2009}. However,
without this read-out later, wave analysis based directly on thresholding
SAC potentials is noisy, making wave front tracking more difficult
and wave 'collisions' more common. In order to compare simulated waves
to recorded waves, Godfrey \emph{et al} 2007 \cite{Godfrey2007} thus
compute a calcium response variable, in which wave fronts are smoothed. 

Our simulations (Figure \ref{fig:realwaves}A), and experiments \cite{Ford2012}
show highly variable participation in wave activity on a cell-cell
basis. Thus to compare our simulations to calcium imaging data, which
is based on filtered RGC activity, we lower the threshold for which
a cell is counted as participating in a wave to the ACh release threshold
(near -55 mV), instead of its spiking threshold. This thus reflects
the `general excitation' of the region, not individual SAC activity,
which makes it better suited for comparison with calcium imaging data.
With this threshold, smoothing our data did not have a significant
effect on wave labeling, so was not implemented (Figure S2).

\subsection*{Bifurcations}

AUTO \cite{Doedel2007a} was used for the numerical continuation of traveling
fronts and bifurcation analysis.

\subsection*{Mathematical analysis}

In order to analyze the model we first perform non-dimensionalisation,
the parameters and scalings are outlined in Table \ref{tab:dimlessparameters}.
Non-dimensional dynamic variables are named in lower-case. To non-dimensionalise
we make the following change of variables: 
\begin{align*}
V=V_{Ca}v,\quad R=r,\quad A=\tau_{ACh}\beta a, & \quad S=\tau_{S}\gamma s
\end{align*}
and scale time (dimensioned quantity here denoted by a capital $T$)
and space (capital $X$) such that 
\begin{align*}
T=\frac{C_{m}}{g_{K}^{M}}t,\quad X=\sqrt{\frac{DC_{m}}{g_{K}^{M}}}x.
\end{align*}
Then:

\textbf{
\begin{align*}
v_{t} & =-\tilde{g}_{Ca}(v)(v-1)-r(v-v_{K})-\tilde{g}_{L}^{M}(v-v_{L})-\tilde{g}_{ACh}(a)(v-v_{ACh}),\\
r_{t} & =\frac{1}{\tilde{\tau}_{R}}[\lambda(v)(r_{\infty}(v)-r)+\tilde{\alpha}s(1-r)],\\
s_{t} & =\frac{1}{\tilde{\tau}_{S}}(g(v)-s),\\
a_{t} & =\nabla^{2}a+\frac{1}{\tilde{\tau}_{ACh}}(g(v)-a),
\end{align*}
}with

\textbf{
\begin{align*}
g(v) & =\frac{1}{1+\exp[-\tilde{\kappa}(v-v_{0})]}\\
\tilde{g}_{Ca}(v) & =\frac{1}{2}\tilde{g}_{Ca}^{M}\left[1+\tanh\left(\frac{v-v_{1}}{v_{2}}\right)\right]\\
\tilde{g}_{ACh}(a) & =\tilde{g}_{ACh}^{M}\frac{\tilde{\delta}a^{2}}{1+\tilde{\delta}a^{2}}\\
\lambda(v) & =\cosh\left(\frac{v-v_{3}}{2v_{4}}\right)\\
r_{\infty}(v) & =\frac{1}{2}\left[1+\tanh\left(\frac{v-v_{3}}{v_{4}}\right)\right].
\end{align*}
}To proceed, let $\epsilon=1/\tilde{\tau}_{R}$, $\epsilon\theta=1/\tilde{\tau}_{S}$
and $\alpha = \pi$\textbf{
\begin{eqnarray*}
f(v,r,a) & = & -\tilde{g}_{Ca}(v)(v-1)-r(v-v_{K})-\tilde{g}_{L}^{M}(v-v_{L})-\tilde{g}_{ACh}(a)(v-v_{ACh}),\\
h(v,r,s) & = & \frac{1}{\tilde{\tau}_{R}}[\lambda(v)(r_{\infty}(v)-r)+\tilde{\alpha}s(1-r)],\\
k(x,v) & = & -x+g(v),
\end{eqnarray*}
}to give:\textbf{
\begin{align}
v_{t} & =f(v,r,a),\nonumber \\
r_{t} & =\epsilon h(v,r,s),\nonumber \\
s_{t} & =\epsilon\theta k(s,v),\nonumber \\
a_{t} & =\nabla^{2}a+\frac{k(a,v)}{\tilde{\tau}_{ACh}},\label{eq:dimlesssystem}
\end{align}
}which we analyze as a fast-slow system as is standard (refer to Ermentrout
and Terman 2010 \cite{Ermentrout2010a}, or Keener and Sneyd 2001 \cite{Sneyd2001}
). The methods described below make no attempt to rigorously establish
the existence of traveling fronts or waves.

\subsection*{Wave fronts}

This scaling of space and time is appropriate in regions where the
diffusion term is relevant, thus the system obtained by letting $\epsilon\to0$\textbf{
\begin{align}
v_{t} & =f(v,a;r),\nonumber \\
a_{t} & =\nabla^{2}a+\frac{k(a,v)}{\tilde{\tau}_{ACh}},\label{eq:fastsystem}
\end{align}
}provides an approximate description of the wave fronts and backs.
The variable $r$ is here considered a parameter.  To construct wave
front solutions in one spatial dimension, change coordinates to a
frame moving with speed $c$:
\[
x'=x-ct,\quad t'=t
\]
so that \textbf{
\begin{align*}
v_{t'}-cv_{x'} & =f(v,a;r),\\
a_{t'}-ca_{x'} & =\nabla^{2}a+\frac{k(a,v)}{\tilde{\tau}_{ACh}}.
\end{align*}
}A traveling front corresponds to the stationary solutions:\textbf{
\begin{align}
0 & =f(v,a;r)+cv'+\epsilon v'',\nonumber \\
0 & =ca'+a''+\frac{k(a,v)}{\tilde{\tau}_{ACh}}.\label{eq:heteroclinic}
\end{align}
}where $'=\frac{d}{dx'}$. A bounded wave front solution, if it exists,
corresponds to a \emph{heteroclinic }orbit connecting the rest fixed
point to the excited fixed point of Equation \ref{eq:heteroclinic}.
In order to obtain this heteroclinic orbit, Equation \ref{eq:fastsystem} was
simulated in MATLAB with a sigmoid function connecting the rest and
excited fixed points for initial data. Once the solution had sufficient
time to converge to the traveling front solution, its profile was
saved, its wave speed calculated, and these two pieces of data put
into the AUTO subpackage homcont to perform continuation
on the system described in Equation \ref{eq:heteroclinic}. This allows
for the computation of $c(r)$ -- the wave speed as a function of
refractory state.

\subsection*{Excitability thresholds}

We seek to determine when the medium is sufficiently excitable to
support a traveling front solution. In two variable neuronal systems
such as Fitzhugh-Nagumo or Morris-Lecar, in which case the fast dynamics
are one dimensional, an energy argument can provide conditions under
which the medium is excitable (\emph{e.g. }Ermentrout and Terman 2010
\cite{Ermentrout2010a}). This is not possible since the fast dynamics
are here two dimensional. So, let $r_{0}$ be the unique fixed point
of Equation \ref{eq:dimlesssystem}. We aim to find parameters for
which $c(r_{0})>0$, which, assuming the front we are studying is
excited to the left and at rest to the right, will correspond to a
propagating front. Parameters for which $c(r_{0})<0$ will generate
a receding front, and thus parameters for which $c(r_{0})=0$ corresponds
to the transition between excitable and not. This is found using AUTO. 

These results are compared to model simulations without the noise
channel. The simulations are performed for different parameters for
a simulated time of 10s, with an initial condition in which the domain
is at rest besides a small cluster of grid points on one side of the
domain. Excitability is determined by measuring if activity above
an 'excited' threshold is observed on the other side of the domain
-- from which we infer that a wave must have traversed the simulated
retina.

\subsection*{Relation between one dimensional analysis and two dimensional simulations}

Our analysis is performed in one spatial dimension only. Moving to
two dimensions has the following effect: the Laplacian in radially
symmetric polar coordinates is
\[
\nabla^{2}=\frac{\partial^{2}f}{\partial r^{2}}+\frac{1}{r}\frac{\partial f}{\partial r},
\]
which, for large $r$, approximates the one dimensional Laplacian
operator. Thus we expect, for waves of large radius, the results we
find for one spatial dimension will apply. Indeed, performing the
above simulations to determine excitability threshold with either
a small cluster of initially excited cells (thus simulating a 2D, but
radially symmetric solution) or with an entire strip of initially
excited cells (thus simulating essentially the 1D dynamics) showed
negligible difference.

\section*{Acknowledgments}

The authors would like to thank Pedro Maia, Eli Shlizerman and Julijana
Gjorgjieva for discussion and feedback on this work.


\section*{Figure Legends}

\clearpage 
\section*{Supporting Information Legends}

\begin{figure}[!ht]
\begin{center}
\includegraphics{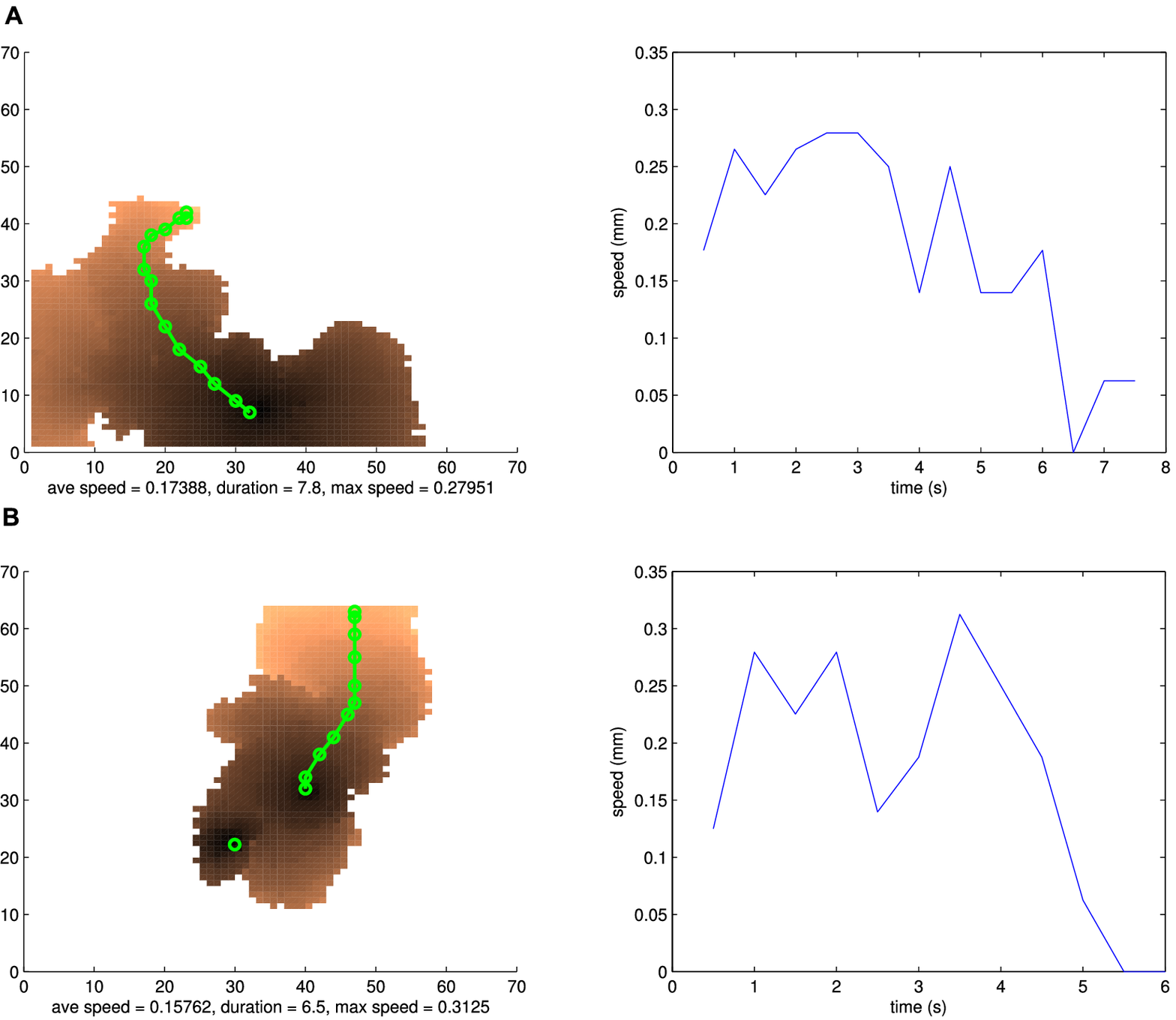}
\end{center}
\caption*{\textbf{Figure S1. Determining speed of wave front for two example
waves.} Black corresponds to earlier times, and orange to later. At
the latest time the wave is active, the wave front which connects the most distal point from
the initiation point is tracked. Starting at
this final point and moving backward in units of 0.5s, the next
closest active point to the current one draws the wave front trajectory in reverse,
as demonstrated in \textbf{A.}. This process does not make sense
if a wave is the result of a collision, as shown in \textbf{B.}. Waves that are involved in a collision have more than one start point (additional start point shown as single green circle), meaning
there is ambiguity in how to apply the wave speed algorithm. The wave speeds for these waves (shown in right subplot) are not counted in our analysis.}
\label{fig:speedcalc}
\end{figure}

\begin{figure}[!ht]
\begin{center}
\includegraphics{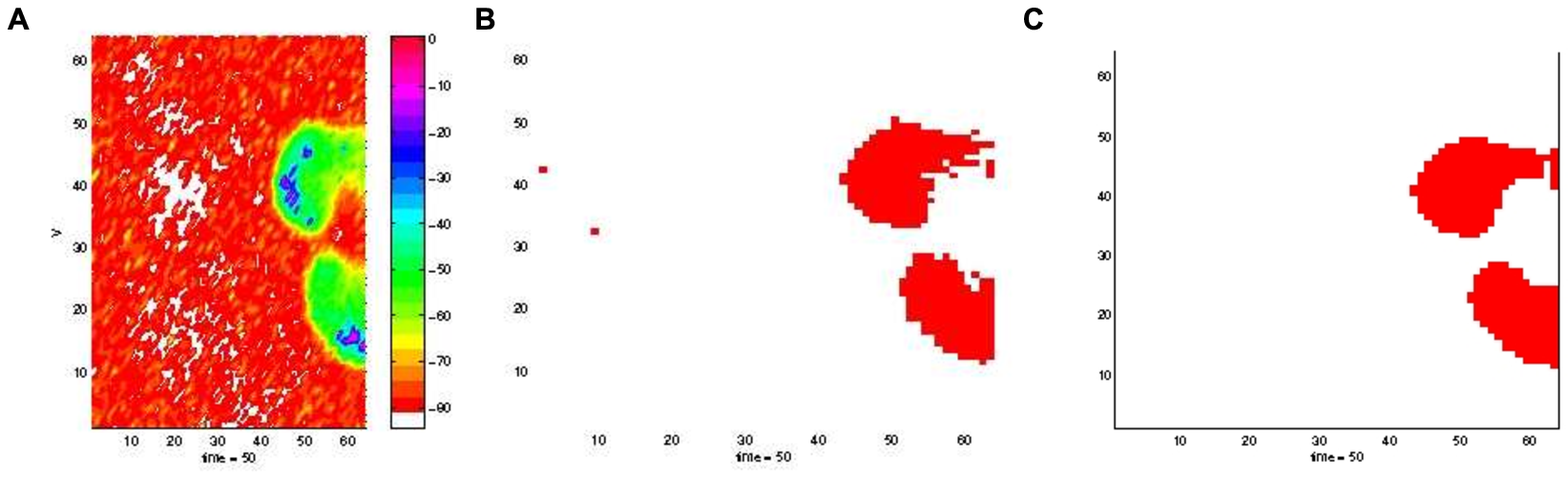}
\end{center}
\caption*{\textbf{Figure S2. Effect of thresholds and smoothing on wave
labeling.} Snapshot of a thresholded and subsequently labeled wave from voltage data \textbf{A.} without \textbf{B.}
and with \textbf{C.} smoothing. Threshold was set to -60 mV to produce wave forms which, by eye, match
waves shown in voltage data.}
\label{fig:smoothing}
\end{figure}

\begin{figure}[!ht]
\begin{center}
\includegraphics{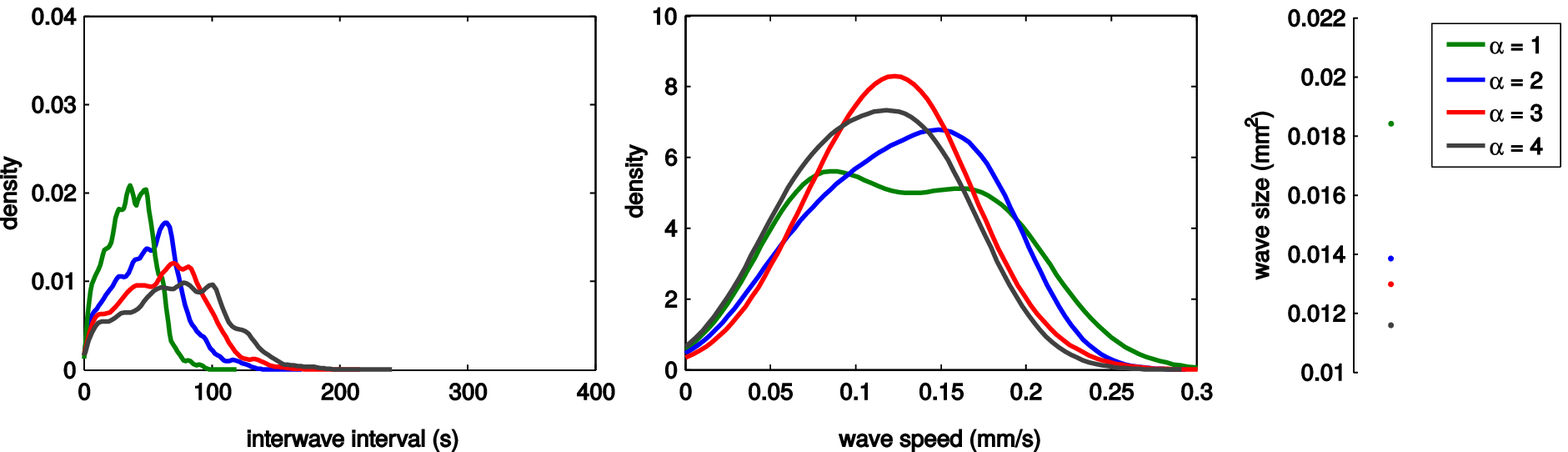}
\end{center}
\caption*{\textbf{Figure S3. Effect of varying sAHP sensitivity parameter $\alpha$.} Statistics following  5000s of simulation of model with indicated values of $\alpha$ -- interwave interval;  
wave speed distribution; and mean wave size. Comparison to Figure \ref{fig:pharma} shows varying $\alpha$ has a similar effect to varying the timescale $\tau_S$. Note that varying the sAHP sensitivity parameter $\alpha$ or varying the rate of activation parameter $\gamma$ have identical impact on the model -- as can be seen from the dimensional analysis summaryized in Table \ref{tab:dimlessparameters}.}
\label{fig:alphaparam}
\end{figure}

\begin{figure}[!ht]
\begin{center}
\includegraphics{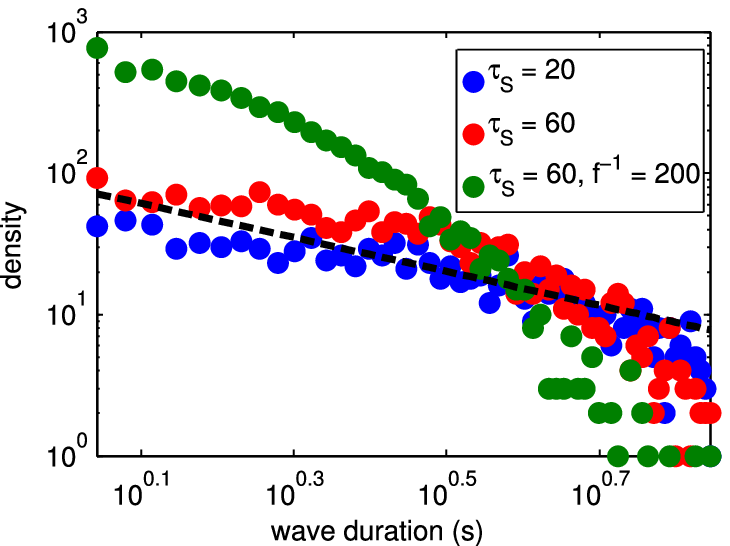}
\end{center}
\caption*{\textbf{Figure S4. Distributions of wave duration (lifetime).}  Distribution of wave duration on logarithmic scale following 5000s simulation on a 128x128 grid point domain for specified values of $\tau_S$ and $f^{-1}$.  For reference, the dotted black curve shows a theoretical distribution of the form $p(s)\sim s^{-\alpha}$ for $\alpha = -1.2$ as observed in large-scale simulations of the DS-FFM model \cite{Pruessner2012}. Comparison to Figure \ref{fig:soc} shows that SOC is not predicted for the parameters drawn in green, which indeed is clearly sub-critical.}
\label{fig:wavedurs}
\end{figure}

\begin{figure}[!ht]
\begin{center}
\end{center}
\caption*{\textbf{Movie S1. 500s of simulation of retinal waves model.} Simulation of 4${}^2$mm of retina played at 10x speed. Left panel to right panel: voltage potential ($V$), sAHP variable ($S$), and acetylcholine variable ($A$). A higher resolution version can be found at https://github.com/benlansdell/retinalwaves}
\end{figure}

\clearpage
\section*{Tables}

\begin{table}[!ht]
\caption{Parameters for retinal waves model}
\begin{tabular}{|c|c|c|c|}
\hline 
Parameter  & Value  & Parameter  & Value \tabularnewline
\hline 
$C_{m}$  & 0.160 nF  & $V_{2}$  & 20 mV  \tabularnewline
\hline 
$V_{Ca}$  & 50 mV  & $V_{3}$  & -25 mV  \tabularnewline
\hline 
$V_{K}$  & -90 mV  & $V_{4}$  & 40 mV  \tabularnewline
\hline 
$V_{syn}$  & 50 mV  & $k$ & 2 \tabularnewline
\hline 
$V_{N}$  & 50 mV  & $\kappa$ & 0.2 mV$^{-1}$ \tabularnewline
\hline 
$V_{L}$  & -70 mV & $V_{0}$ & -40 mV \tabularnewline
\hline 
$g_{ACh}^{M}$  & 2 nS  & $D$ & 0.01 mm$^{2}$s$^{-1}$  \tabularnewline
\hline 
$g_{Ca}^{M}$  & 10 nS  & $\beta$  & 5 nM$\cdot$s$^{-1}$  \tabularnewline
\hline 
$g_{K}^{M}$  & 30 nS  & $\delta$  & 800 nM$^{-2}$ \tabularnewline
\hline 
$g_{L}^{M}$ & 3 nS & $\tau_{R}$  & 5 s  \tabularnewline
\hline 
$V_{1}$  & -20 mV  & $\tau_{ACh}$ & 0.2 s \tabularnewline
\hline 
$V_{n}$ & 50mV & $\tau_{S}$ & 60 s \tabularnewline
\hline 
$\alpha$ & 2 & $\mu$ & 5 \tabularnewline
\hline 
$L$  & 2 mm  & $\gamma$ & 0.3 s$^{-1}$ \tabularnewline
\hline 
\end{tabular}
\begin{flushleft}\textbf{Dimensioned parameters used in numerical simulations unless otherwise specified in text.} 
The chosen length of domain (2mm) and a simulation of a 64$\times$64 square lattice corresponds to approximation one
SAC per grid point assuming an average distance between SACs of approximately 30 microns.
\end{flushleft}
\label{tab:parameters}
\end{table}

\begin{table}[!ht]
\caption{Dimensionless parameters for retinal waves model}
\begin{tabular}{|c|c|c|c|c|c|}
\hline 
Parameter  & Relation  & Value  & Parameter  & Relation  & Value\tabularnewline
\hline 
$v_{K}$  & $V_{K}/V_{Ca}$  & -1.8  & $v_{0}$  & $V_{0}/V_{Ca}$  & -0.8 \tabularnewline
\hline 
$v_{ACh}$  & $V_{ACh}/V_{Ca}$  & 1 & $v_{1}$  & $V_{1}/V_{Ca}$  & -0.4 \tabularnewline
\hline 
$v_{L}$  & $V_{L}/V_{Ca}$  & -1.4  & $v_{2}$ & $V_{2}/V_{Ca}$ & 0.4 \tabularnewline
\hline 
$\bar{g}{}_{Ca}$  & $g_{Ca}^{M}/g_{K}^{M}$  & 1/3  & $v_{3}$  & $V_{3}/V_{Ca}$  & -0.5 \tabularnewline
\hline 
$\bar{g}{}_{ACh}$  & $g_{ACh}^{M}/g_{K}^{M}$  & 1/15  & $v_{4}$  & $V_{4}/V_{Ca}$  & 0.8 \tabularnewline
\hline 
$\bar{g}{}_{L}$ & $g_{L}^{M}/g_{K}^{M}$ & 1/10  & $\tilde{\kappa}$ & $\kappa V_{Ca}$  & 10\tabularnewline
\hline 
$\frac{1}{\tilde{\tau}_{R}}$ & $C_{m}/(\tau_{R}g_{K}^{M})$  & 0.001067  & $\tilde{D}$ & $(C_{m}D)/(g_{K}^{M}L^{2})$  & 1.333$\times10^{-5}$\tabularnewline
\hline 
$\frac{1}{\tilde{\tau}_{ACh}}$ & $C_{m}/(\tau_{ACh}g_{K}^{M})$  & 0.02667  & $\tilde{\delta}$  & $\delta\tau_{ACh}^{2}\beta^{2}$ & 800\tabularnewline
\hline 
$\frac{1}{\tilde{\tau}_{S}}$ & $C_{m}/(\tau_{S}g_{K}^{M})$ & 8.88$\times10^{-5}$ & $\tilde{\alpha}$ & $\alpha\tau_{S}\gamma$ & 36\tabularnewline
\hline 
\end{tabular}
\begin{flushleft}\textbf{Dimensionless parameters used in analysis of model and their values.} 
\end{flushleft}
\label{tab:dimlessparameters} 
\end{table}

\end{document}